\documentclass{aastex631}

\newcommand{\kmsMpc}{\ensuremath{\mbox{km s}^{-1} \,\mbox{Mpc}^{-1}}}

\begin{document}

\title{A dark siren measurement of the Hubble constant with the LIGO/Virgo gravitational wave event GW190412 and DESI galaxies}

\shorttitle{Dark siren measurement of LIGO/Virgo GW190412 with DESI}
\shortauthors{Ballard, Palmese, et al.}

\author{W. Ballard}
\affiliation{McWilliams Center for Cosmology, Carnegie Mellon University,
5000 Forbes Avenue, Pittsburgh, PA 15213, USA}
\author[0000-0002-6011-0530]{A. Palmese}\thanks{apalmese@andrew.cmu.edu}
\affiliation{McWilliams Center for Cosmology, Carnegie Mellon University,
5000 Forbes Avenue, Pittsburgh, PA 15213, USA}
\author[0000-0003-2362-0459]{I. Maga\~na~Hernandez}
\affiliation{McWilliams Center for Cosmology, Carnegie Mellon University,
5000 Forbes Avenue, Pittsburgh, PA 15213, USA}

\author[0000-0001-5537-4710]{S.~BenZvi}
\affiliation{Department of Physics \& Astronomy, University of Rochester, 206 Bausch and Lomb Hall, P.O. Box 270171, Rochester, NY 14627-0171, USA}

\author{J.~Moon}
\affiliation{Department of Physics and Astronomy, Sejong University, Seoul, 143-747, Korea}

\author{A.~J.~Ross}
\affiliation{Center for Cosmology and AstroParticle Physics, The Ohio State University, 191 West Woodruff Avenue, Columbus, OH 43210, USA}
\affiliation{Department of Astronomy, The Ohio State University, 4055 McPherson Laboratory, 140 W 18th Avenue, Columbus, OH 43210, USA}

\author{G.~Rossi}
\affiliation{Department of Physics and Astronomy, Sejong University, Seoul, 143-747, Korea}

\author{J.~Aguilar}
\affiliation{Lawrence Berkeley National Laboratory, 1 Cyclotron Road, Berkeley, CA 94720, USA}

\author[0000-0001-6098-7247]{S.~Ahlen}
\affiliation{Physics Dept., Boston University, 590 Commonwealth Avenue, Boston, MA 02215, USA}

\author[0000-0002-8622-4237]{R.~Blum}
\affiliation{NSF's NOIRLab, 950 N. Cherry Ave., Tucson, AZ 85719, USA}

\author{D.~Brooks}
\affiliation{Department of Physics \& Astronomy, University College London, Gower Street, London, WC1E 6BT, UK}

\author{T.~Claybaugh}
\affiliation{Lawrence Berkeley National Laboratory, 1 Cyclotron Road, Berkeley, CA 94720, USA}

\author[0000-0002-1769-1640]{A.~de la Macorra}
\affiliation{Instituto de F\'{\i}sica, Universidad Nacional Aut\'{o}noma de M\'{e}xico,  Cd. de M\'{e}xico  C.P. 04510,  M\'{e}xico}

\author[0000-0002-4928-4003]{A.~Dey}
\affiliation{Department of Physics \& Astronomy, University College London, Gower Street, London, WC1E 6BT, UK}

\author{P.~Doel}
\affiliation{Department of Physics \& Astronomy, University College London, Gower Street, London, WC1E 6BT, UK}

\author[0000-0002-2890-3725]{J.~E.~Forero-Romero}
\affiliation{Departamento de F\'isica, Universidad de los Andes, Cra. 1 No. 18A-10, Edificio Ip, CP 111711, Bogot\'a, Colombia}
\affiliation{Observatorio Astron\'omico, Universidad de los Andes, Cra. 1 No. 18A-10, Edificio H, CP 111711 Bogot\'a, Colombia}

\author[0000-0003-3142-233X]{S.~Gontcho A Gontcho}
\affiliation{Lawrence Berkeley National Laboratory, 1 Cyclotron Road, Berkeley, CA 94720, USA}

\author{K.~Honscheid}
\affiliation{Center for Cosmology and AstroParticle Physics, The Ohio State University, 191 West Woodruff Avenue, Columbus, OH 43210, USA}
\affiliation{Department of Physics, The Ohio State University, 191 West Woodruff Avenue, Columbus, OH 43210, USA}

\author[0000-0001-6356-7424]{A.~Kremin}
\affiliation{Lawrence Berkeley National Laboratory, 1 Cyclotron Road, Berkeley, CA 94720, USA}

\author[0000-0003-4962-8934]{M.~Manera}
\affiliation{Departament de F\'{i}sica, Serra H\'{u}nter, Universitat Aut\`{o}noma de Barcelona, 08193 Bellaterra (Barcelona), Spain}
\affiliation{Institut de F\'{i}sica da Altes Energies (IFAE), The Barcelona Institute of Science and Technology, Campus UAB, 08193 Bellaterra Barcelona, Spain}

\author[0000-0002-1125-7384]{A.~Meisner}
\affiliation{NSF's NOIRLab, 950 N. Cherry Ave., Tucson, AZ 85719, USA}

\author{R.~Miquel}
\affiliation{Instituci\'{o} Catalana de Recerca i Estudis Avan\c{c}ats, Passeig de Llu\'{\i}s Companys, 23, 08010 Barcelona, Spain}
\affiliation{Institut de F\'{i}sica da Altes Energies (IFAE), The Barcelona Institute of Science and Technology, Campus UAB, 08193 Bellaterra Barcelona, Spain}

\author[0000-0002-2733-4559]{J.~Moustakas}
\affiliation{Department of Physics and Astronomy, Siena College, 515 Loudon Road, Loudonville, NY 12211, USA}

\author[0000-0001-7145-8674]{F.~Prada}
\affiliation{Instituto de Astrof\'{i}sica de Andaluc\'{i}a (CSIC), Glorieta de la Astronom\'{i}a, s/n, E-18008 Granada, Spain}

\author[0000-0002-9646-8198]{E.~Sanchez}
\affiliation{CIEMAT, Avenida Complutense 40, E-28040 Madrid, Spain}

\author[0000-0003-1704-0781]{G.~Tarl\'{e}}
\affiliation{University of Michigan, Ann Arbor, MI 48109, USA}

\author[0000-0002-4135-0977]{Z.~Zhou}
\affiliation{National Astronomical Observatories, Chinese Academy of Sciences, A20 Datun Rd., Chaoyang District, Beijing, 100012, P.R. China}

\collaboration{1000}{(DESI Collaboration)}

\begin{abstract}

We present a measurement of the Hubble Constant $H_0$ using the gravitational wave event GW190412, an asymmetric binary black hole merger detected by LIGO/Virgo, as a dark standard siren. This event does not have an electromagnetic counterpart, so we use the statistical standard siren method and marginalize over potential host galaxies from the Dark Energy Spectroscopic Instrument (DESI) survey. 
GW190412 is well-localized to 12 deg$^2$ (90\% credible interval), so it is promising for a dark siren analysis. The dark siren value for 
$H_0=85.4_{-33.9}^{+29.1}$ \kmsMpc, 
with a posterior shape that is consistent with redshift overdensities. When combined with the bright standard siren measurement from GW170817 we recover $H_0=77.96_{-5.03}^{+23.0}$ \kmsMpc, consistent with both early and late-time Universe measurements of $H_0$. This work represents the first standard siren analysis performed with DESI data, and includes the most complete spectroscopic sample used in a dark siren analysis to date.

\end{abstract}


\section{Introduction} \label{sec:intro}
A promising application of gravitational waves (GW) is in the measurement of cosmological parameters such as the Hubble Constant $H_0$. The GW signals from binary black hole mergers can be used as standard sirens \citep{Schutz_1986a} due to the mergers being absolute distance indicators. 
If we are unable to find an electromagnetic counterpart and pinpoint the host galaxy to measure a redshift, we can use a statistical approach taking advantage of measurements of the redshift distribution of populations of potential host galaxies \citep{2023AJ....166...22G}. 
Novel measurements of $H_0$ are important to understand the origin of the ``Hubble Tension'' (\citealt{Riess_2022}; \citealt{2020}). 

In this note, we use the galaxy catalog from the Dark Energy Spectroscopic Instrument (DESI; \citealt{DESI:2022xcl,DESI:2023dwi}) to perform a dark siren measurement of $H_0$. Compared to photometric redshifts used in previous works \citep{Palmese_2023}, the spectroscopic redshifts measured by DESI are $\sim2$ orders of magnitude more accurate, allowing for a tighter constraint on $H_0$. Compared to other spectroscopic catalogs used in dark siren analyses (e.g.  \citealt{LIGOScientific:2021aug}), DESI offers a significantly more complete sample of galaxies. Moreover, we have carried out the first dedicated spectroscopic observations of a dark siren.

We focus on GW190412, which resulted from the merger of 30 $M_{\odot}$ and 6 $M_{\odot}$ black holes and had no electromagnetic counterpart \citep{Abbott_2020}. This event has one of the smallest localizations ($\sim 12~\mbox{deg}^2$) of all GW detections so far, making it ideal to carry out a dark siren measurement. Its localization falls within the DESI footprint. 

\section{Data} \label{sec:data}

The GW data used come from the GW190412 LIGO/Virgo skymap \citep{Abbott_2020}. Each skymap pixel includes both a spatial probability and information for a luminosity distance $d_L$ likelihood. 

For the galaxies, we use the DESI Iron internal data release from the first year of DESI operations, along with the daily reductions, which include all targets observed to date. We focus on the BGS (Bright Galaxy Sample; \citealt{Hahn_2023}), containing roughly 10 million galaxies out to $z\sim0.5$. 
We consider galaxies with $r$-band absolute magnitude brighter than $-21.5$ 
 and reliable redshift estimations.
 We also limit our analysis to galaxies within the 90\% CI area given by the GW skymap, and to redshifts consistent with the GW $d_L$ given our $H_0$ prior. 
After the appropriate cuts are made, 6039 galaxies from Iron are analyzed. We use large scale structure (LSS) weights to account for variations in survey completeness due to various selection effects \citep{DESI:2023bgx,DESI:2023ytc}.


The ``daily'' reductions sample includes the final DESI BGS observations in this part of the sky, as well as a secondary targeting program performed around GW190412 \citep{DESI:2023ytc}. 
The sample, which contains 8442 galaxies, allows us to investigate a more complete set of galaxies without LSS weights and to estimate the effect of using different versions of the spectroscopic pipeline.

\section{Results} \label{sec:results}

To estimate $H_0$, we follow a similar approach to \cite{Chen_2018,Palmese_2020}, assuming a uniform prior between $[20,140]$ \kmsMpc. We assume a flat $\Lambda$CDM cosmology with $\Omega_m=0.3$, and uncertainties are the $68\%$ credible interval (CI). The analysis is blinded to account for potential confirmation bias. The blinding is removed once we are confident in the quality cuts made for our dataset.

\begin{figure}
    \centering
    \includegraphics[scale=0.5]{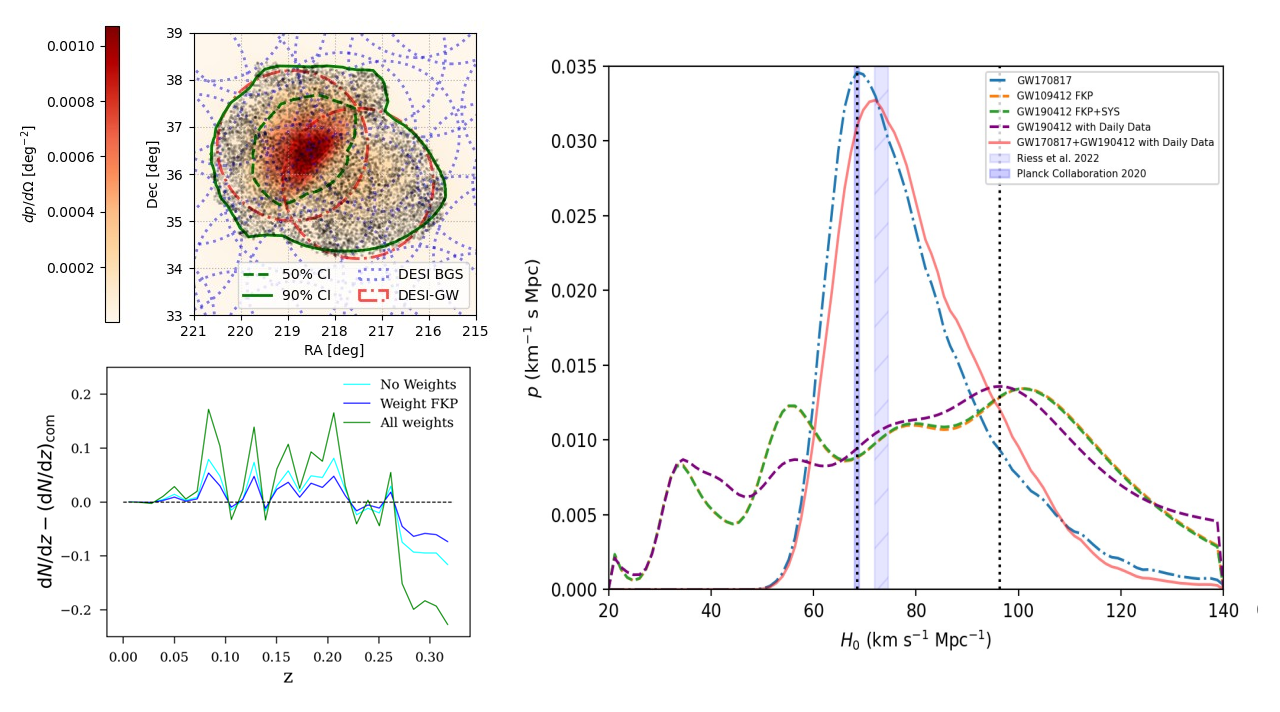}
    \caption{\emph{Upper left:} GW190412 skymap. The spatial probability is given by the color map and the 50\% and 90\% CI are shown by the green dashed and solid lines. Each black point is a DESI galaxy. The blue circles represent the bright-time DESI pointings in the region. The red circles are the DESI pointings taken as part of secondary target GW follow-up program. 
    \emph{Bottom left:} DESI galaxies' redshift distribution subtracted by that of a uniform-in-comoving-volume distribution, to highlight the presence of overdensities along the GW190412 line-of-sight.  \emph{Right:} $H_0$ posteriors from this analysis. 1$\sigma$ constraints from \cite{Riess_2022} and \cite{2020} are also shown. The vertical lines to indicate the 68\% CI of our combined GW170817+GW190412 fiducial result.}
    \label{fig1}

\end{figure}

The posterior distribution of $H_0$ using GW190412 as a dark siren is shown in Figure \ref{fig1}. The dashed lines show the impact of using different sets of weights (or no weights). Multiple peaks in the posterior are observed, corresponding to several redshift overdensities in the GW localization volume. The width of the peaks are expected to be broad, since we are marginalizing over thousands of GW host galaxies and there is a degree of uncertainty associated with $d_L$ and, to a smaller degree, with the  redshifts. 
We note that the peaks are narrower than those found in an analysis that uses photometric redshifts with the same event \citep{Palmese_2023}, due to the smaller redshift uncertainty. 

We explore the impact of each LSS weight by applying them one by one. The `FKP' weights, which reweight volume elements based on the expected tracer's number density, are expected to have the most significant impact on the posterior, smoothing out some of the lowest redshift peaks, where observations are more likely to be complete already from earlier passes. `WEIGHT\_SYS' takes into account fluctuations in the imaging data that impact the target selection. Because we do not consider a large angular scale, 
we do not see any significant impact on the posterior due to this weight. `WEIGHT\_COMP' accounts for the completeness of target observations due to fiber assignment. As a cross-check, we note that using observations from the full ``daily'' sample provides very similar results to the Y1 Iron data using FKP weights.

For our fiducial analysis using the daily reductions we find $H_0$ to be 
$H_0=85.4_{-33.9}^{+29.1}$
\kmsMpc (median likelihood and 68\% CI). When combined with the GW170817 posterior \citep{10.1093/mnras/staa1120}, we find 
$H_0=77.96_{-9.48}^{+18.4}$ \kmsMpc. This result is broadly consistent with the results from both Planck and R22. 
We plan to use additional gravitational wave events as dark sirens alongside the available DESI data for a more constraining analysis.

\section{Acknowledments}\label{sec:acknowledgments}
\begin{acknowledgments}
This material is based upon work supported by NSF Grant No. 2308193, the U.S. Department of Energy (DOE), Office of Science, Office of High-Energy Physics, under Contract No. DE–AC02–05CH11231, and by the National Energy Research Scientific Computing Center, a DOE Office of Science User Facility under the same contract. Additional support for DESI was provided by the U.S. National Science Foundation (NSF), Division of Astronomical Sciences under Contract No. AST-0950945 to the NSF’s National Optical-Infrared Astronomy Research Laboratory; the Science and Technology Facilities Council of the United Kingdom; the Gordon and Betty Moore Foundation; the Heising-Simons Foundation; the French Alternative Energies and Atomic Energy Commission (CEA); the National Council of Science and Technology of Mexico (CONACYT); the Ministry of Science and Innovation of Spain (MICINN), and by the DESI Member Institutions: \url{https://www.desi.lbl.gov/collaborating-institutions}. Any opinions, findings, and conclusions or recommendations expressed in this material are those of the author(s) and do not necessarily reflect the views of the U. S. National Science Foundation, the U. S. Department of Energy, or any of the listed funding agencies.

The authors are honored to be permitted to conduct scientific research on Iolkam Du’ag (Kitt Peak), a mountain with particular significance to the Tohono O’odham Nation.
\end{acknowledgments}


\bibliographystyle{}
\bibliography{sample631}{}
\bibliographystyle{aasjournal}

\end{document}